\documentclass[aps,prl,twocolumn,showpacs,floatfix]{revtex4}
\usepackage{graphicx}
\usepackage{times}
\usepackage{nicefrac}
\usepackage{amsmath}
\usepackage{amsfonts}
\usepackage{amssymb}
\usepackage{amsthm}
\usepackage{epsf}
\usepackage{bm}
\usepackage{bbm}

\usepackage{dcolumn}
\newcolumntype{.}{D{x}{}{-1}}

\newcommand{\be}{\begin{eqnarray}}
\newcommand{\ee}{\end{eqnarray}}
\newcommand{\la}{\langle}
\newcommand{\ra}{\rangle}

\newcommand{\veps}{\varepsilon}

\newcommand{\pr}{\prime}

\newcommand{\bfr}{{\bf r}}

\newcommand{\rg}{{\rm g}}
\newcommand{\aZ}{\alpha Z}

\begin{document}

\title{Nuclear polarization study: New frontiers for tests of QED in heavy highly charged ions}

\author{Andrey V. Volotka$^{1,2}$ and G\"unter Plunien$^{1}$}

\affiliation{
$^1$ Institut f\"ur Theoretische Physik, Technische Universit\"at Dresden,
Mommsenstra{\ss}e 13, D-01062 Dresden, Germany \\
$^2$ Department of Physics, St. Petersburg State University,
Oulianovskaya 1, Petrodvorets, 198504 St. Petersburg, Russia \\
}

\begin{abstract}
A systematic investigation of the nuclear-polarization effects in one- and few-electron
heavy ions is presented. The nuclear-polarization corrections in the zeroth and first
orders in $1/Z$ are evaluated to the binding energies, the hyperfine splitting, and the
bound-electron g factor. It is shown, that the nuclear-polarization contributions can
be substantially canceled simultaneously with the rigid nuclear corrections. This allows
for new prospects for probing the QED effects in strong electromagnetic field and the
determination of fundamental constants.
\end{abstract}


\maketitle
%
%
The enormous progress made in experimental investigations of heavy highly charged
ions during the last decades (see, e.g., Refs.~\cite{gumberidze:2007:87,
beiersdorfer:2010:074032,sturm:2013:620,noertershaeuser:2013:014016,lochmann:2014:xxx}
and references therein) has triggered the vigorous development of {\it ab initio}
QED theory in the presence of strong nuclear fields. The relativistic behaviour
of electrons in highly charged ions requires a fully relativistic description
from the very beginning, i.e. nonperturbative in the $\aZ$ parameter, where
$Z$ is the nuclear charge number. This plays a key role in contrast to
QED theory for light atomic systems, where the parameter $\aZ$ is employed
as an expansion parameter. Over the last decades an essential progress has
been achieved in theoretical calculations of various spectroscopic
properties of highly charged ions, such as transition energies, hyperfine
splitting (HFS), and g factor (see Refs.~\cite{sapirstein:2008:25,shabaev:2011:60,
volotka:2013:636} for reviews). In many cases further improvement of the
achieved theoretical accuracy seems strongly limited by the lack of the knowledge
of the nuclear properties. E.g., in the case of the g factor of H-like lead
ion the uncertainty of nuclear charge distribution correction is the main source
of the total uncertainty, and in the case of the HFS in the H-like bismuth ion
the uncertainty of the nuclear magnetization distribution correction (so-called
Bohr-Weisskopf effect) strongly masks the QED contributions.
In Ref.~\cite{shabaev:2001:3959} it was proposed to consider a specific difference
of the HFS values of H- and Li-like ions with the same nucleus, where the
uncertainty of the Bohr-Weisskopf effect is significantly reduced and the QED
effects can be tested on the level of a few percent. In the case of the g factor
similar cancellations of the finite nuclear size corrections have been recognized
for the specific differences of the g factors of H- and Li-like ions in
Ref.~\cite{shabaev:2002:062104} and of H- and B-like ions in
Ref.~\cite{shabaev:2006:253002}, respectively. These differences can be calculated
with a substantially higher accuracy, which opens excellent perspectives for a test
of the QED effects and even provides a possibility for an independent determination
of the fine structure constant from the strong-field QED theory.

Another nuclear effect appears due to the intrinsic nuclear dynamics, where
the nucleus interacting with electrons via the radiation field can undergo real or
virtual electromagnetic excitations. The latter effect leads to the nuclear-polarization
(NP) correction, e.g., to the binding energy of the electrons. Being restricted to
phenomenological descriptions of the nucleon-nucleon interaction the NP correction
sets the ultimate accuracy limit up to which QED corrections can be tested
in highly charged ions. Therefore, an important question should be addressed:
To which extent can NP corrections be canceled in specific differences? In this Letter,
we rigorously examined the NP and the screened NP corrections to the binding energies,
HFS, and g factor of heavy highly charged ions. We analyze the ratio of the finite
nuclear size and the NP corrections and consequently evaluate the NP contribution to
the specific differences, designed for the cancellation of the finite nuclear size
and the Bohr-Weisskopf effect.

In Refs.~\cite{plunien:1989:5428,plunien:1991:5853} a relativistic field theoretical
approach to the NP corrections in electronic atoms incorporating the effects due to
virtual collective nuclear excitations within the framework of bound-state QED for
atomic electrons was developed. This approach was successfully applied in calculations
of NP corrections to the binding energies \cite{plunien:1989:5428,plunien:1991:5853,
labzowsky:1994:371,plunien:1995:1119,nefiodov:1996:227}, to the HFS \cite{nefiodov:2003:35},
and to the bound-electron g factor \cite{nefiodov:2002:081802}.

\begin{figure}
\includegraphics[width=0.2\textwidth]{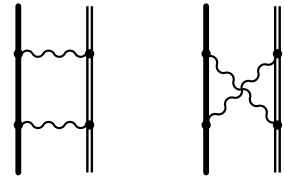}
\caption{Feynman diagrams representing the lowest-order nuclear-polarization effect to
the electron binding energy. The bound electron (double line) interacts with the nucleus
(heavy line) in its ground state via the exchange of virtual photons (wavy lines).}
\label{fig:np}
\end{figure}

%
The lowest order diagrams describing the NP effect are depicted in Fig.~\ref{fig:np}.
As virtual nuclear excitations we account for the dominant ones arising from
the collective nuclear dynamics, such as rotations of deformed nuclei, harmonic
surface vibrations, and giant resonances. Since the velocities associated with
the collective nuclear dynamics are nonrelativistic we can restrict to the
nuclear charge-density fluctuation (electric multipole transitions) and neglect
contributions arising from fluctuations of the nuclear vector current (magnetic
multipole transitions). It is most suitable to employ Coulomb gauge and keeping
the longitudinal component ${\tilde D}_{00}$ of effective photon propagator
${\tilde D}_{\mu\nu}$ only. It describes the interaction between the electrons
and virtual nuclear transitions and takes the form \cite{plunien:1989:5428,
plunien:1991:5853}:
\be
{\tilde D}_{00}(\bfr_1,\bfr_2,\omega) &=& \sum_{L M} B(EL;L \rightarrow 0)
           \frac{2\omega_L}{\omega^2-\omega_L^2+i0}\nonumber\\
 &\times& F_L(r_1)F_L(r_2)\, Y_{L M}(\Omega_1) Y^\ast_{L M}(\Omega_2)\,.
\ee
Here $\omega_L$ is the nuclear excitation energy and $B(EL;L \rightarrow 0)$ is
corresponding reduced electric transition probability. This form of the
propagator is also very suitable for numerical calculations, since it 
exclusively depends on phenomenological quantities such as transition energies
$\omega_L$, and corresponding electric transition strengths.
The radial dependence carried by the functions $F_L$ may be specified utilizing,
e.g., a sharp surface model for describing the corresponding collective nuclear
multipole transition densities \cite{plunien:1989:5428,plunien:1991:5853,
labzowsky:1994:371,plunien:1995:1119,nefiodov:1996:227}. The nuclear ground-state
sphere radius $R_0$ is determined as
$R_0 = \sqrt{5/3}\,\la r^2 \ra^{1/2}$, where $\la r^2 \ra^{1/2}$ is the
root-mean-square charge radius.

The energy shift due to the lowest-order NP effect is given by
\be
\Delta E_{\rm NP} = e^2 \frac{i}{2\pi}\int_{-\infty}^\infty d\omega
                    \sum_n \frac{\la a n | {\tilde D}_{00}(\omega) | n a \ra}{\veps_a-\omega-\veps_n u}\,,
\ee
where the summation runs over the complete Dirac spectrum, and $u = 1 - i0$
preserves the proper treatment of poles of the electron propagator. In
Table~\ref{tab:np} the leading order NP corrections are presented for the
$1s$, $2s$, and $2p_{1/2}$ binding energies in ${}^{208}_{82}$Pb and
${}^{238}_{92}$U ions.
\begin{table}
\caption{Nuclear-polarization $\Delta E_{\rm NP}$ and finite nuclear size
$\Delta E_{\rm FNS}$ corrections to the $1s$, $2s$, and $2p_{1/2}$ binding
energies in ${}^{208}_{82}$Pb and ${}^{238}_{92}$U ions. The nuclear-polarization
corrections are compared with the previous calculations \cite{nefiodov:1996:227}.
The corresponding ratios of the nuclear-polarization and finite nuclear size
contributions $\Delta_{\rm NP/FNS}$ are presented.}
\label{tab:np}
\tabcolsep5pt
\begin{tabular}{lr@{}lr@{}lr@{}l} \hline\hline
State   & \multicolumn{2}{c}{$1s$}
                      & \multicolumn{2}{c}{$2s$}
                                   & \multicolumn{2}{c}{$2p_{1/2}$}
                                               \\[1mm]\hline\\[-2mm]
\multicolumn{7}{c}{${}^{208}_{82}$Pb nucleus}  \\[1mm]
$\Delta E_{\rm NP}$ (meV)
        &  -28&.89    &  -5&.033   & -0&.4249  \\
        &  -29&.3$^a$ &  -5&.0$^a$ &           \\
        &  -31&.8$^b$ &  -5&.5$^b$ &           \\
$\Delta E_{\rm FNS}$ (eV)
        &   67&.18    &  11&.66    &  0&.9991  \\
$\Delta_{\rm NP/FNS}$ ($10^{-3}$)
        &   -0&.430   &  -0&.431   & -0&.425   \\[1mm]
\multicolumn{7}{c}{${}^{238}_{92}$U nucleus}   \\[1mm]
$\Delta E_{\rm NP}$ (meV)
        & -188&.2     & -35&.88    & -4&.153   \\
        & -197&.6$^a$ & -37&.2$^a$ & -4&.2$^a$ \\
        & -213&.4$^b$ & -40&.9$^b$ & -4&.6$^b$ \\
$\Delta E_{\rm FNS}$ (eV)
        &  198&.6     &  37&.73    &  4&.412   \\
$\Delta_{\rm NP/FNS}$ ($10^{-3}$)
        &   -0&.947   &  -0&.951   & -0&.941   \\
\hline\hline
\end{tabular}
\\
$^a$Ref.~\cite{nefiodov:1996:227}: direct numerical integration (a).\\
$^b$Ref.~\cite{nefiodov:1996:227}: B-spline calculations (b).
\end{table}
For the low-lying rotational and vibrational nuclear excitations the experimental
values for the excitation energies $\omega_L$ and electric transition strengths
$B(EL;L \rightarrow 0)$ are taken from Ref.~\cite{208Pb:2007} for the
${}^{208}_{82}$Pb ion and from  Ref.~\cite{238U:2002} for the ${}^{238}_{92}$U ion.
The corresponding $\omega_L$ and $B(EL;L \rightarrow 0)$ values for the giant
resonances have been estimated employing the phenomenological energy-weighted
sum rules \cite{rinker:1978:397}. The summation over the spectrum of the Dirac
equation has been performed employing the dual-kinetic-balance finite basis set
method \cite{shabaev:2004:130405} with basis functions constructed from B splines
\cite{sapirstein:1996:5213}. The Dirac spectrum is calculated in the field of extended
nuclei utilizing nuclear charge density distributions with recent values for the radii
$\la r^2 \ra^{1/2} = 5.5010$ fm and $\la r^2 \ra^{1/2} = 5.8569$ fm in the case of
lead and uranium ions, respectively. As one can see from Table~\ref{tab:np} the
obtained results are in a fair agreement with the previous calculations presented
in Ref.~\cite{nefiodov:1996:227}, and a better agreement is found with the results
obtained by the direct numerical integration.

In Table~\ref{tab:np} we also present the corresponding leading order finite nuclear
size corrections $\Delta E_{\rm FNS}$ together with a ratio of the NP and finite
nuclear size terms $\Delta_{\rm NP/FNS}$ defined as
$\Delta_{\rm NP/FNS} = \Delta E_{\rm NP}/\Delta E_{\rm FNS}$. These ratios appear
to behave rather similar for all the considered electron state. This means that
canceling the finite nuclear size corrections in an energy difference the NP effect
will be also reduced to a large extent.

\begin{figure}
\includegraphics[width=0.45\textwidth]{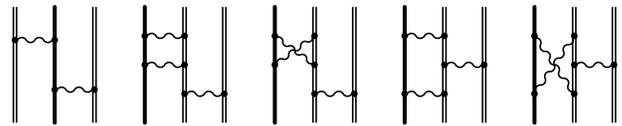}
\caption{Feynman diagrams representing the screened nuclear-polarization effect to the
electron level energy. Notations are the same as in Fig.~\ref{fig:np}.}
\label{fig:snp}
\end{figure}

However, the hydrogenic excited energy states are not always accessible experimentally,
e.g., at present, the highest accuracy was achieved in measurements of the $2p_{1/2}-2s$
transition energy in heavy Li-like $^{238}$U$^{89+}$ ion \cite{beiersdorfer:2005:233003}.
Therefore, it becomes of distinct importance to investigate the NP effect in few-electron
ions. In the presence of other electrons, in addition to the leading order one-electron
NP correction, terms combining the interelectronic-interaction and NP effects appear.
In analogy to the corresponding QED corrections we refer to them as the screened NP
contribution. The diagrams of the NP corrections to the one-photon exchange are depicted
in Fig.~\ref{fig:snp}.
Expressions for the energy shift due to this effect can be derived according to the
Feynman diagrams depicted but are to lengthy to be presented here. Their evaluations
have been performed in both Feynman and Coulomb gauges for the photon propagator
describing the electron-electron interaction, thus providing an accurate check of the
numerical procedure. The results obtained for the screened NP corrections to the
$(1s)^2$ binding energy and to the $(1s)^2 2s$, $(1s)^2 2p_{1/2}$, and $(1s)^2 2p_{3/2}$
ionization energies in ${}^{208}_{82}$Pb and ${}^{238}_{92}$U ions are presented in
Table~\ref{tab:snp}.
\begin{table}
\caption{Screened nuclear-polarization and screened finite nuclear size corrections
$\Delta E_{\rm SNP}$ and $\Delta E_{\rm SFNS}$, respectively, to the $(1s)^2$
binding energy and to the $(1s)^2 2s$, $(1s)^2 2p_{1/2}$, and $(1s)^2 2p_{3/2}$
ionization energies (with the opposite sign) in ${}^{208}_{82}$Pb and
${}^{238}_{92}$U ions. The ratio of the screened nuclear-polarization and
screened finite nuclear size contributions $\Delta_{\rm SNP/SFNS}$ are
presented.}
\label{tab:snp}
\tabcolsep3pt
\begin{tabular}{lr@{}lr@{}lr@{}lr@{}l} \hline\hline
State   & \multicolumn{2}{c}{$(1s)^2$}
                   & \multicolumn{2}{c}{$(1s)^2 2s$}
                              & \multicolumn{2}{c}{$(1s)^2 2p_{1/2}$}
                                         & \multicolumn{2}{c}{$(1s)^2 2p_{3/2}$}
                                                    \\[1mm]\hline\\[-2mm]
\multicolumn{9}{c}{${}^{208}_{82}$Pb nucleus}       \\[1mm]
$\Delta E_{\rm SNP}$ (meV)
        &  0&.8441 &  0&.3017 &  0&.1218 &  0&.0335 \\
$\Delta E_{\rm SFNS}$ (eV)
        & -1&.9629 & -0&.6988 & -0&.2763 & -0&.0863 \\
$\Delta_{\rm SNP/SFNS}$ ($10^{-3}$)
        & -0&.430  & -0&.432  & -0&.441  & -0&.388  \\[1mm]
\multicolumn{9}{c}{${}^{238}_{92}$U nucleus}        \\[1mm]
$\Delta E_{\rm SNP}$ (meV)
        &  5&.498  &  2&.048  &  0&.9369 &  0&.1793 \\
$\Delta E_{\rm SFNS}$ (eV)
        & -5&.802  & -2&.152  & -0&.9816 & -0&.2117 \\
$\Delta_{\rm SNP/SFNS}$ ($10^{-3}$)
        & -0&.948  & -0&.952  & -0&.955  & -0&.847  \\
\hline\hline
\end{tabular}
\end{table}
As one can see, the screened NP correction is comparable
with the leading order term, and has to be taken into account in specific
differences constructed for eliminating the finite nuclear size effect.
In Table~\ref{tab:snp} we also present the finite nuclear size effect coming
from the first-order interelectronic-interaction correction, the so-called
screened finite nuclear size correction $\Delta E_{\rm SFNS}$, together
with the corresponding ratio $\Delta_{\rm SNP/SFNS}$ defined as
$\Delta_{\rm SNP/SFNS} = \Delta E_{\rm SNP} / \Delta E_{\rm SFNS}$.
The ratio of screened NP and finite nuclear size corrections
appears to be rather stable for different electronic configurations. This
opens a possibility to eliminate in such differences not only the finite
nuclear size corrections, but also the NP terms to
a rather large extent.

As an example for such a cancellation, let us consider the following difference.
One of the most precise measurements were performed for the $1s$ Lamb shift
$\Delta E^{(1s)}$ in H-like uranium U$^{91+}$ \cite{gumberidze:2005:223001} and
for the $2p_{1/2}-2s$ transition energy $\Delta E^{(2p_{1/2}-2s)}$ in U$^{89+}$
\cite{beiersdorfer:2005:233003}. In both cases the uncertainty of the finite
nuclear size correction essentially contributes to the total theoretical
error bars. Thus, we can construct the following specific difference
$\Delta' E = \Delta E^{(2p_{1/2}-2s)} + \xi \Delta E^{(1s)} \approx 355.8$ eV,
where the parameter $\xi = 0.161856$ is chosen in a way to cancel the leading
order and the screened finite nuclear size terms. Such a cancellation is also
rather stable with respect to the employed nuclear charge distribution model.
The NP corrections are canceled in $\Delta' E$ up to $10^{-4}$ eV
opening a possibility for unprecedented tests of strong-field QED.
Although we have considered here only nuclear excitations of electric type,
this conclusion will hold in general, since the cancellation being discussed
is observed for each individual nuclear excitation. In view of this, we can
expect, that this cancellation will be rather independent from the employed
nuclear models.

%
Let us now turn to the NP effects to the HFS in few-electron ions. The HFS transition
line in Li-like Bi$^{80+}$ has been recently observed and directly measured in a laser
spectroscopy measurement at GSI \cite{noertershaeuser:2013:014016,lochmann:2014:xxx}.
This has allowed for the first time to compare experimental and theoretical values for
the specific difference between HFS of H- and Li-like bismuth ions. Although the present
experimental accuracy is smaller than the theoretical one \cite{volotka:2012:073001,
andreev:2012:022510} in the future SPECTRAP Penning trap facility this will be improved
by three orders of magnitude \cite{andelkovic:2013:033423}. A substantial progress in
the theoretical calculations of the specific difference \cite{volotka:2009:033005,
glazov:2010:062112,volotka:2012:073001,andreev:2012:022510} allows to improve the
theoretical accuracy by an order of magnitude, and the present uncertainty is partially
restricted by the NP correction, which was so far known only for the $1s$ HFS
\cite{nefiodov:2003:35}.

\begin{figure}
\includegraphics[width=0.45\textwidth]{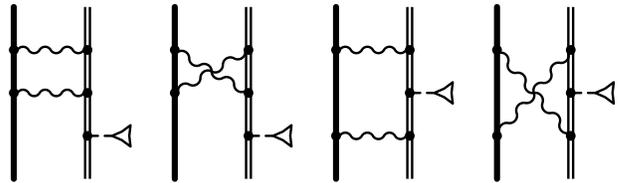}
\caption{Feynman diagrams representing the lowest-order nuclear-polarization effect in
the presence of an external potential. The dashed line terminated with the triangle
denotes the interaction with the external magnetic field. Notations are the same as
in Fig.~\ref{fig:np}.}
\label{fig:np:g}
\end{figure}

\begin{figure}
\includegraphics[width=0.45\textwidth]{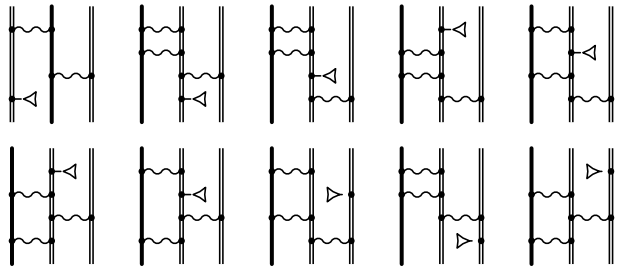}
\caption{Feynman diagrams representing the screened nuclear-polarization effect in
the presence of an external potential. For brevity, we depict here only the direct
part of the nuclear-polarization correction for all the diagrams except the first
one. Notations are the same as in Fig.~\ref{fig:np}.}
\label{fig:snp:g}
\end{figure}

In this Letter we present for the first time results for the leading order and the
screened NP corrections to the HFS of H-, Li-, and B-like bismuth ions. The leading
order and the screened NP effect are given by the diagrams depicted in
Figs.~\ref{fig:np:g} and \ref{fig:snp:g}, respectively.

The basic expressions for the leading order diagrams are similar to those derived in
Ref.~\cite{nefiodov:2003:35}, and for the screened diagrams they are rather bulky and will
be presented elsewhere. The numerical procedure has been accurately checked by employing the
Feynman and Coulomb gauges for the photon propagator describing the interelectronic
interaction. For the nuclear parameters of low-lying vibrational levels of the nearly
spherical odd-even ${}^{209}_{83}$Bi nucleus we have employed the corresponding collective
vibration levels in the neighboring even-even isotope of ${}^{208}_{82}$Pb (weak coupling limit).
Moreover, we have evaluated the effect of single-nucleon excitations and the effect going
beyond the weak-coupling limit. Both have been found to be negligible in the case of
${}^{209}_{83}$Bi nucleus compared to the effect of collective core excitations.
The detailed consideration will be presented elsewhere. The obtained results for the leading
order and the screened NP corrections $\Delta E_{\rm HFS,NP}$ and $\Delta E_{\rm HFS,SNP}$,
respectively, are presented in Table~\ref{tab:hfs}.
\begin{table}
\caption{Individual contributions to the leading order $\Delta E_{\rm HFS,NP}$
and the screened $\Delta E_{\rm HFS,SNP}$ nuclear-polarization corrections to
the ground state hyperfine splitting of H-, Li, and B-like ${}^{209}_{83}$Bi ions
in $\mu$eV. The total nuclear-polarization contribution to the specific differences
$\Delta^\pr_{\rm HL} E_{\rm HFS,NP}$ and $\Delta^\pr_{\rm HB} E_{\rm HFS,NP}$
are also presented in $\mu$eV.}
\label{tab:hfs}
\begin{tabular}{llll|cc} \hline\hline
State                    &           $1s$  &           $2s$ &     $2p_{1/2}$
                      & $\Delta^\pr_{\rm HL} E_{\rm HFS,NP}$ & $\Delta^\pr_{\rm HB} E_{\rm HFS,NP}$ \\ \hline
$\Delta E_{\rm HFS,NP}$  &\phantom{-}50.34 &\phantom{-}8.865&\phantom{-}0.730&           &          \\
                         &\phantom{-}55$^a$&                &                &           &          \\
$\Delta E_{\rm HFS,SNP}$ &                 &          -0.340&          -0.095& 0.025(12) & -0.09(5) \\
\hline\hline
\end{tabular}
\\
$^a$Ref.~\cite{nefiodov:2003:35}.\\
\end{table}
Here, we employ the nuclear single-particle model for the description of the Bohr-Weisskopf
effect. In the case of the $1s$ state a fair agreement has been achieved with the previous
value \cite{nefiodov:2003:35}, that was obtained within the point-like magnetic moment
approximation. Now we introduce two specific differences: the first is between the HFS of
H- and Li-like ions $\Delta^\pr_{\rm HL} E_{\rm HFS}$ defined as
$\Delta^\pr_{\rm HL} E_{\rm HFS} = \Delta E^{[(1s)^2 2s]}_{\rm HFS}
- \xi_{\rm HL} \Delta E^{[1s]}_{\rm HFS}$, and the second is between the HFS of H- and
B-like ions $\Delta^\pr_{\rm HB} E_{\rm HFS} = \Delta E^{[(1s)^2 (2s)^2 2p_{1/2}]}_{\rm HFS}
- \xi_{\rm HB} \Delta E^{[1s]}_{\rm HFS}$. The parameters $\xi_{\rm HL}$ and
$\xi_{\rm HB}$ are chosen in a way to cancel the Bohr-Weisskopf effects, and they
appear to be rather stable with respect to variations of the nuclear model of the
magnetization distribution \cite{shabaev:2001:3959}. In the case of ${}^{209}_{83}$Bi
these parameters are chosen to be $\xi_{\rm HL} = 0.16886$ and $\xi_{\rm HB} = 0.014459$.
In Table~\ref{tab:hfs} we present the obtained results for the total NP contribution
to the specific differences under consideration. As we can stress now the
NP effects are essentially reduced in the both differences. The obtained results
for the $\Delta^\pr_{\rm HL} E_{\rm HFS,NP}$ and $\Delta^\pr_{\rm HB} E_{\rm HFS,NP}$
appear to be rather stable with respect to the changes of the charge and magnetization
distribution models. In view of this we assign an uncertainty of 50\% to the total NP
corrections to the specific differences. This result opens the possibility for further
theoretical improvements of the HFS specific differences and tests of the magnetic sector
of strong-field QED. Moreover, this can lead to the determination of the nuclear magnetic
moments from a comparison between the theoretical and experimental values for the specific
differences.

%
We can now also consider the situation for the bound-electron g factor. The leading
order and the screened NP corrections are given by the same diagrams as for the HFS,
see, Figs.~\ref{fig:np:g} and \ref{fig:snp:g}, respectively. In Table~\ref{tab:g} we
present our numerical results obtained for the g factors of H-, Li-, and B-like lead ions.
\begin{table}
\caption{Individual contributions to the leading order $\Delta$g$_{\rm NP}$ and
the screened $\Delta$g$_{\rm SNP}$ nuclear-polarization corrections to the ground
state g factor of H-, Li, and B-like ${}^{208}_{82}$Pb ions in $10^{-8}$. The
total nuclear-polarization contribution to the specific differences
$\Delta$g$^\pr_{\rm HL,NP}$ and $\Delta$g$^\pr_{\rm HB,NP}$ are also
presented in $10^{-8}$. The leading order nuclear-polarization corrections
are compared with the previous calculations \cite{nefiodov:2002:081802}.}
\label{tab:g}
\tabcolsep5pt
\begin{tabular}{llll|ll} \hline\hline
State                 & $1s$  &  $2s$   & $2p_{1/2}$ & $\Delta$g$^\pr_{\rm HL,NP}$
                                                   & $\Delta$g$^\pr_{\rm HB,NP}$ \\ \hline
$\Delta$g$_{\rm NP}$  &-19.77 & -3.444  &-0.291             &           &              \\
                      &-22$^a$& -3.8$^a$&-0.32$^a$          &           &              \\
$\Delta$g$_{\rm SNP}$ & &\phantom{-}0.129 &\phantom{-}0.104 & -0.013(6) & 0.006(3)     \\
                      &       &         &                   &           & 0.004(6)$^b$ \\
\hline\hline
\end{tabular}
\\
$^a$Ref.~\cite{nefiodov:2002:081802}; $^b$Ref.~\cite{shabaev:2006:253002}.\\
\end{table}
Our values for $\Delta$g$_{\rm NP}$ are in a reasonable agreement with the previous
calculations of Ref.~\cite{nefiodov:2002:081802}. We also present the results for the
specific differences between the H- and Li-like g factors $\rg^\pr_{\rm HL}$ defined by
$\rg^\pr_{\rm HL} = \rg^{[(1s)^2 2s]} - \xi_{\rm HL}\,\rg^{[1s]}$, and between the H-
and B-like g factors
$\rg^\pr_{\rm HB} = \rg^{[(1s)^2 (2s)^2 2p_{1/2}]} - \xi_{\rm HB}\,\rg^{[1s]}$.
In the case of Pb ions the $\xi$ parameters are chosen to be $\xi_{\rm HL} = 0.1670264$
\cite{shabaev:2002:062104} and $\xi_{\rm HB} = 0.0097416$ \cite{shabaev:2006:253002}.
As one can see from the table the NP corrections are canceled in a specific differences
by about two orders of magnitude. Our conservative estimation of the total uncertainty
is of about 50\% of the effect. In comparison with the rough estimation of the screened
NP correction made in Ref.~\cite{shabaev:2006:253002}, here we rigorously take into
account the first-order interelectronic-interaction correction to the nuclear polarization.
We have also evaluated the effect originating from nuclear excitations induced via
via two types of magnetic interactions: one is due to the interaction with a constant
magnetic field, and the other one is due to the magnetic interaction with an electron.
In Ref.~\cite{jentschura:2006:102} such corrections were referred to as nuclear magnetic
susceptibility corrections to the g factor. The contributions of this effect to the
g factor are found to be $-6.6 \times 10^{-11}$, $-1.1 \times 10^{-11}$, and
$-0.4 \times 10^{-11}$ for the $1s$, $2s$, and $2p_{1/2}$ states, respectively. The
corresponding contributions to the specific differences are at least by an order of
magnitude smaller than our uncertainty. Thus, we can state an improvement of the accuracy
of NP correction to the specific difference $\rg^\pr_{\rm HB}$ by a factor of 2.
In view of this new result we can push back the ultimate limit, defined by the NP
effects, to the specific difference $\rg^\pr_{\rm HB}$ and consequently to the possible
accuracy of the determination of the fine structure constant. The recommended value
of the fine structure constant according to the recent CODATA \cite{mohr:2012:1527}
is $\alpha = 1/137.035\,999\,074(44)$. The corresponding uncertainty in the specific
difference $\delta \rg^\pr_{\rm HB}[\alpha] = 5 \times 10^{-11}$ is 1.5 times larger
than the theoretical limit given by the NP uncertainty
$\delta \rg^\pr_{\rm HB, NP} = 3 \times 10^{-11}$. Another principal uncertainty in
the specific difference is coming from the remaining finite nuclear size effect.
However, this uncertainty can be substantially reduced by employing the more accurate
charge distribution parameters obtained from muonic atoms \cite{kozhedub:2008:032501,
zatorski:2012:063005}.

%
To conclude, we have evaluated the leading order and the screened NP corrections to
the binding energies, HFS, and bound-electron g factor of heavy highly charged ions. The interelectronic-interaction
effects have been rigorously evaluated within the QED perturbation theory up to the
first order in $1/Z$. The effect of the nuclear polarization has been evaluated for
the specific differences constructed in a way to cancel the nuclear size corrections.
In all cases considered here it turns out, that the NP corrections determining the
ultimate accuracy cancel substantially. Therefore, the rigorous investigations of the
specific differences provide a unique opportunity to test the strong-field QED with much
higher accuracy than expected before. The ultimate accuracy of the nuclear
polarization for the specific difference $\rg^\pr_{\rm HB}$ has been now improved by
a factor of 2. This may clear the way for more accurate determination of the fine
structure constant from the strong-field QED with a precision similar to the one obtained
from the investigations of the free-electron g factor.

%
Valuable conversations with D. A. Glazov, A. V. Nefiodov, K. Pachucki,
and V. M. Shabaev are gratefully acknowledged.
The work reported in this paper was supported by DFG (Grant No. VO 1707/1-2).
A.V.V. is grateful to the Mainz Institute for Theoretical Physics (MITP)
for its hospitality and support.

\end{document}